\documentclass{article}

\usepackage[dvips]{color}
\usepackage{epsfig}
\usepackage{amsfonts,amssymb,latexsym,amsmath,amscd}
\usepackage{dcolumn}
\usepackage{times,mathptm,color}

\usepackage{anysize}
\marginsize{1in}{1in}{1in}{1in}

\definecolor{gold}{rgb}{0.85,.66,0}
\definecolor{plum}{rgb}{0.45,0,.66}
\definecolor{bgn}{rgb}{0,.85,.46}  
\definecolor{myred}{rgb}{1,.1,.16}  

\newcommand{\bl}{\begin{itemize}}
\newcommand{\el}{\end{itemize}}

\newcommand{\INC}{{\tt INC}}
\newcommand{\CINC}{{\tt CINC}}
\newcommand{\CNOT}{{\tt CNOT}}

    {
    \smallskip
    \refstepcounter{theorem}
    \noindent
    {\bf Example \arabic{section}.\arabic{theorem}} \ \ }
    {\hspace*{\fill}{$\Diamond$}
    \smallskip}

    {
    \smallskip
    \refstepcounter{theorem}
    \noindent
    {\bf Remark \arabic{section}.\arabic{theorem}} \ \ }
    {\hspace*{\fill}{$\Diamond$}
    \smallskip}

    {
    \smallskip
    \refstepcounter{theorem}
    \noindent
    {\bf Remark \Alph{section}.\arabic{theorem}} \ \ }
    {\hspace*{\fill}{$\Diamond$}
    \smallskip}

    {
    \smallskip
    \refstepcounter{theorem}
    \noindent
    {\bf Definition \arabic{section}.\arabic{theorem}} \ \ }
    {\hspace*{\fill}{\ }
    \smallskip}

    {
    \smallskip
    \refstepcounter{theorem}
    \noindent
    {\bf Definition \Alph{section}.\arabic{theorem}} \ \ }
    {\hspace*{\fill}{\ }
    \smallskip}

\newenvironment{prf}[1][]
    {
    \noindent
    {\bf Proof{#1}:  }
    }
    {\hspace*{\fill}{$\Box$}\smallskip}

    {
    \noindent
    {\bf Sketch{#1}:  }
    }
    {\hspace*{\fill}{$\Box$}\smallskip}

    {
    \noindent
    {\bf Reference:  }
    }
    {\hspace*{\fill}{$\odot$}\smallskip}

\newtheorem{theorem}{Theorem}[section]
\newtheorem{proposition}[theorem]{Proposition}
\newtheorem{lemma}[theorem]{Lemma}
\newtheorem{definition}[theorem]{Definition}

\include{Qcircuit}

\newcommand{\controlu}{*-=[][F]{\phantom{\bullet}}}

\title{Efficient Circuits for Exact-Universal Computation With Qudits}

\author{Gavin~K.~Brennen$^1$, Stephen~S.~Bullock$^{2,3}$ and
Dianne~P.~O'Leary$^{2,4}$}

\date{September 22, 2005}

\begin{document}

\maketitle

\smallskip

\noindent
$^1${\small Atomic Physics Division, National Institute of Standards and Technology, Gaithersburg, MD 20899-8420}

\noindent
$^2${\small Mathematical and Computational Sciences Division, National Institute of Standards and Technology}

{\small Gaithersburg, MD 20899-8910}

\noindent
$^3${\small Center for Computing Sciences, Institute for Defense Analyses, Bowie, MD  20715-4300}

\noindent
$^4${\small Department of Computer Science and Institute for Advanced Computer Studies, University of Maryland,
College Park, MD 20742.}

\begin{abstract}
This paper concerns the efficient implementation of quantum 
circuits for qudits.
We show that controlled two-qudit gates can be implemented
without ancillas and prove that the gate library containing
arbitrary local unitaries and one two-qudit gate, $\CINC$,
is exact-universal.
A recent paper [S.Bullock, D.O'Leary, and G.K. Brennen, Phys. Rev. Lett. 
{\bf 94}, 230502 (2005)] describes quantum circuits for qudits which
require $O(d^n)$ two-qudit gates for state synthesis and $O(d^{2n})$ 
two-qudit gates for unitary synthesis, matching the respective lower 
bound complexities.
In this work, we present the state synthesis 
circuit in much greater detail and prove that it is correct.  
Also, the $\lceil (n-2)/(d-2) \rceil$ ancillas required in
the original algorithm may be removed without changing the asymptotics.
Further, we present a new algorithm for unitary synthesis, inspired by 
the QR matrix decomposition, which is also asymptotically optimal.   
\end{abstract}

\section{Introduction}
\label{sec:intro}

A qudit is a $d$-level generalization of a qubit, i.e. the
one-qudit Hilbert space splits orthogonally as
\begin{equation}
\mathcal{H}(1,d) \ = \ 
\mathbb{C}\{ \ket{0} \} \oplus \mathbb{C} \{ \ket{1} \} \oplus \cdots
\oplus \mathbb{C} \{ \ket{d-1} \}
\end{equation}
while the $n$-qudit state-space is 
$\mathcal{H}(n,d)=[\mathcal{H}(1,d)]^{\otimes n}$.
Thus for $N=d^n$, closed-system evolutions of $n$ qudits
are modeled by $N \times N$ unitary matrices.  Qudit circuit diagrams
then factor such unitaries into two-qudit operations
$I_{d^{n-2}} \otimes V$ where $V$ is a $d^2 \times d^2$ unitary
matrix, or more generally into
similarity transforms of such gates by particle-swaps.
The algorithmic complexity of an evolution may then be thought of
as the number of two-qudit gates required to build it.
A degree of freedom argument \cite{Knill_state} leads one to
guess that exponentially many gates are required for most unitary
evolutions, since the space of all $N \times N$ unitary matrices
is $d^{2n}$-dimensional.  Indeed, this space of evolutions is a manifold
so the argument may be made rigorous using smooth topology,
and thus $\Omega(d^{2n})$ gates are required for exact-universality.
Yet until quite recently the best qudit circuits contained
$O(n^2d^{2n})$ gates \cite{MuthukrishnanStroud:00}.  In contrast,
$O(4^n)$ gates were known to suffice for qubits ($d=2$) \cite{Vartiainen:04}, 
presenting the possibility that qudits are genuinely less efficient for
$d$ not a power of two.

Quite recently, an explicit $O(d^{2n})$ construction was
achieved \cite{sharpqudit}.  It uses the spectral decomposition
of the unitary matrix desired and also a new \emph{state synthesis
circuit} \cite{Deutsch_state,Knill_state,Shende_state,Bergholm_state}.
Given a $\ket{\psi} \in \mathcal{H}(n,d)$, a state-synthesis circuit
for $\ket{\psi}$ realizes some unitary $U$ such that $U\ket{\psi}=\ket{0}$.
There are $2d^n-2$ real degrees of freedom in a normalized state ket 
$\ket{\psi}$, which may
be used to prove that circuits for generic states cost
$\Omega(d^n)$ two-qudit gates.  This is in sharp contrast to the
case of classical logic, where $O(n)$ inverters may produce any
bit-string.  The most recent qudit state-synthesis circuit
\cite{sharpqudit}
contains $(d^n-1)/(d-1)$ two-qudit
gates, and in fact each is a singly-controlled one-qudit operator
$\wedge_1(V)=I_{d^{2}-d} \oplus V$.

There are two ways to employ an asymptotically optimal
state synthesis circuit in order to obtain asymptotically optimal
unitary circuits.  The first is to exploit the spectral decomposition,
which involves a three part circuit for each 
eigenstate of the unitary:   
building an eigenstate \cite{Knill_state,sharpqudit}, applying a conditional
phase to one logical basis ket, and unbuilding the eigenstate.
We here introduce a second option, the {\bf Triangle} algorithm,
which uses the 
state-synth circuit with extra controls to reduce the
unitary to upper triangular form.
Recursive counts of the number of control
boxes show that it is also asymptotically optimal (Cf.  
\cite{Vartiainen:04}.)  
Although these algorithms are unlikely to be used to implement
general unitary matrices, they can be usefully applied
to improving subblocks of larger circuits (peephole optimization).

Finally, this work also addresses two further topics in which
qudit circuits lag behind qubit circuits.  First, to date the smallest 
gate library for exact universality with qudits uses arbitrary 
locals complemented 
by a continuous one parameter two-qudit gate \cite{selectionQR}.  In contrast,
it is well known \cite{Deutsch_state} that any computation on qubits 
can be realized using gates from the 
library $\{U(2)^{\otimes n},\CNOT \}$.  We prove that the library 
$\{U(d)^{\otimes n},\CINC \}$,
where $\CINC$ is the qudit generalization of the $\CNOT$ gate,
is exactly universal.  Second, the
first asymptotically optimal qubit quantum circuit exploited
a single ancilla qubit \cite{Vartiainen:04} and current
constructions require none \cite{Bergholm_state,Shende_state}, while 
qudit diagrams tend to suppose $\lceil (n-2)/(d-2) \rceil$ ancilla
qudits.  Here we present methods which
realize a $k$-controlled operation 
$\wedge_{k}(V)=I_{d^{k+1}-d}\oplus V$ in $O[(k+2)^{2+\log_2 d}]$ gates 
without the need for any ancilla.  This makes all qudit asymptotics
competitive with their qubit counterparts.  However, it is not known
whether the Cosine-Sine Decomposition (CSD) is useful for building
qudit circuits, despite the fact that all best-practice qubit
exact universal circuits exploit this matrix decomposition.

The paper is organized as follows.  \S \ref{sec:one_control} improves
on earlier constructions of $\wedge_1(V)$ gates, which are ubiquitous
in later sections.  \S \ref{sec:new_control} presents a new circuit
for a qudit $\wedge_k(V)$ gate which are later
used to produce the first $O(d^{2n})$ gate unitary circuits without ancilla.
\S \ref{sec:algorithm} details the recent state synthesis algorithm
as an iteration over a new $\clubsuit$-sequence and exploits the new
constructions to prove it is correct.  \S \ref{sec:triangle} presents
a new asymptotically optimal unitary circuit inspired by the
$QR$ matrix factorization and compares it with a previous algorithm
based on spectral decomposition.
\S \ref{sec:applics} discusses two applications of the state
synthesis algorithm

\section{Notation and conventions}

The Hilbert spaces
$\mathcal{H}(1,d)$ and $\mathcal{H}(n,d)$ are defined in the
introduction.  On $\mathcal{H}(1,d)$,
the inverter for bits has two important generalizations for dits:
\begin{equation}
\begin{array}{lcl}
\sigma_x \oplus I_{d-2} \ket{j} & = & 
\left\{ \begin{array}{rr}
\ket{1} & j=0 \\
\ket{0} & j=1 \\
\ket{j} & 2 \leq j \leq d-1 \\
\end{array}
\right.
\\
\\
\INC \ket{j}  & = & \ket{(j+1) \mbox{ mod } d} \\
\end{array}
\end{equation}
We use the latter symbol rather than the more typical $X$ since
this operation is a modular increment.
This leads to two generalizations of the quantum controlled-not,
$\wedge_1(\sigma_x \oplus I_{d-2})$ and $\wedge_1(\INC)=\CINC$.
The usual symbol for a controlled-not when appearing in a qudit
circuit diagram refers to \CINC.  Controls
represented by a black bubble in qudit circuit diagrams fire
on control state $\ket{d-1}$.  

As new notation, the $\clubsuit$-sequence is introduced in \S
\ref{sec:clubsuit}.  This plays a role analogous to the Gray code
in earlier $d=2$ constructions and is a particular sequence of words of
$n$-letters.  Although these words might themselves be called
sequences, we prefer to call an individual word (e.g.
$1100\clubsuit \clubsuit$) a \emph{term} and reserve
``sequence'' exclusively for the $\clubsuit$-sequence of $(d^n-1)/(d-1)$ terms.

\section{Optimizing singly-controlled one-qudit unitaries}
\label{sec:one_control}

Several operators $\wedge_1(V)$ appear in later circuits.
Thus, it is worthwhile to optimize this computation in our gate
libraries.  For qubits, CNOT-optimal circuits for $\wedge_1(V)$
are known \cite{SongKlappenecker}.  The qudit case is open.
Here we improve the $\wedge_1(V)$ circuit in that work and further
prove for the first time that
$U(d)^{\otimes n} \sqcup \{ \CINC, \CINC^{-1}\}$ is exact-universal.
Since $\CINC^{-1}=\CINC^{d-1}$, this also demonstrates that
$U(d)^{\otimes n} \sqcup \{\CINC\}$ is exact-universal.
This is a smaller universal library than that presented in earlier
work \cite{selectionQR}.

Thus, consider the question of factoring $\wedge_1(V)$.
Let $\{\ket{\psi_k}\}_{k=0}^{d-1}$ be the eigenkets of $V$ with
eigenvalues $\{\mbox{e}^{i \theta_k}\}_{k=0}^{d-1}$.  Let
$W_k$ be some one-qudit unitary with $W_k\ket{0}=\ket{\psi_k}$,
e.g. the appropriate one-qudit Householder reflection 
(See \S \ref{sec:house}.)
Finally, let $\Phi_k$ be a controlled one-qudit phase unitary
given by
$\Phi_k = \wedge_1[ I_d + (\mbox{e}^{i \theta_k} - 1) \ket{0}\bra{0}]$.
Then note that
$V=\prod_{k=0}^{d-1} W_k [ I_d + (\mbox{e}^{i \theta_k} - 1) \ket{0}\bra{0}]
W_k^\dagger$.  Thus 
$\wedge_1(V)$ can be implemented by the following circuit:
\begin{equation}
\label{eq:one-control-circuit}
\Qcircuit @R 0.75em @C 0.75em {
& \ctrl{1} & \qw & & & &  
\qw & \multigate{1}{\Phi_0} & \qw & 
\qw & \multigate{1}{\Phi_1} & \qw & \push{\cdots} & 
\qw & \multigate{1}{\Phi_{d-1}} & \qw & \qw & \\
& \gate{V} & \qw & & \ustick{\cong} & 
& \gate{W_0^\dagger} & \ghost{\Phi_0} & \gate{W_0} 
& \gate{W_1^\dagger} & \ghost{\Phi_1} & \gate{W_1} & \push{\cdots}
& \gate{W_{d-1}^\dagger} & \ghost{\Phi_{d-1}} & \gate{W_{d-1}} & \qw 
}
\end{equation}
Thus, we have reduced the question to building $\Phi_k$
in terms of $U(d)^{\otimes n}$ and $\CINC$.

Building $\Phi_k$ requires some preliminary remarks.  Suppose
we have $\xi \in \mathbb{C}$, $| \xi |=1$.  Consider the diagonal
unitary of the corresponding geometric sequence:
$D = \sum_{j=0}^{d-1} \xi^j \ket{j}\bra{j}$.  Recall that \INC \ is
the increment permutation, i.e.
\INC $\ket{j} = \ket{(j+1)\mbox{ mod }d}$.  Thus permuting the diagonal
entries, $\INC \; D \; \INC^{-1} \; = \; \xi^{d-1}\ket{0}\bra{0}+
\sum_{j=1}^{d-1} \xi^{j-1} \ket{j}\bra{j}$.  Hence
\begin{equation}
\INC \; D \; \INC^{-1}\;  D^{-1} \ = \ 
\xi^{d-1} \ket{0} \bra{0} + \xi^{-1} \sum_{j=1}^{d-1} \ket{j} \bra{j} 
\ = \ (\xi^{-1}I_d)\big( \xi^d\ket{0}\bra{0}+
\sum_{j=1}^{d-1} \ket{j}\bra{j} \big).
\end{equation} 
Now generalizing a standard trick from qubits, note further
that 
\begin{equation}
\wedge_1(\xi I_d)=\big( \sum_{j=0}^{d-2} \ket{j} \bra{j}
+ \xi \ket{d-1}\bra{d-1} \big) \otimes I_d,
\end{equation}
so that a controlled global-phase
is in fact a local operation.  Hence taking
$\xi=\mbox{e}^{i \theta_k/d}$, we obtain in particular an expression
for $\Phi_k$ of Equation \ref{eq:one-control-circuit} in terms of
\CINC \ and \CINC$^{-1}$:
\begin{equation}
\begin{array}{lcl}
\Phi_k & = & \wedge_1(\xi I_d) \; \CINC\; (I_d \otimes D) \; 
\CINC^{-1} \; (I_d \otimes D^{-1}) \\
& = &
\big[ \big( \sum_{j=0}^{d-2} \ket{j} \bra{j}
+ \xi \ket{d-1}\bra{d-1} \big) \otimes I_d \big]
\; \CINC \; (I_d \otimes D) \; \CINC^{-1} \; (I_d \otimes D^{-1}).
\end{array}
\end{equation}
Hence, $\wedge_1(V)$ may be realized using gates from $U(d)^{\otimes n}$
along with $d$ copies of \CINC \ and $d$ copies of
\CINC$^{-1}$.

Recall that these circuits may be expanded into circuits in terms
of $\wedge_1(\sigma_x \oplus I_{d-2})$.  Indeed, when viewed as permutations,
\INC \ and \INC$^{-1}$ factor into $d$ flips. To see this, 
consider  $0 \leq j < k
\leq d-1$ and let $(j k)$ denote the flip permutation
$j \leftrightarrow k$ of $\{0,1,\ldots,d-1\}$. Then
\begin{equation}
\label{eq:lots_flips}
\INC \ = \ (0 1) \circ (1 2) \circ \cdots \circ (d-2 \; d-1).
\end{equation}
Since $\wedge_1[(j k)]$ is equivalent to $\wedge_1(\sigma_x \oplus I_{d-2})$
up to permutations within $U(d)^{\otimes n}$, we see that \CINC \ 
and \CINC$^{-1}$ may be implemented using $d-1$ copies of the
controlled-flip.  Thus, $\wedge_1(V)$ may also be realized using
$2d(d-1)$ copies of the $\wedge_1(\sigma_x \oplus I_{d-2})$ gate.

\noindent
{\bf Remark:}
Note that the controlled-flip is also equivalent to
$\wedge_1(I_{d-2} \oplus \sigma_z)$,
making blockwise use of the $2 \times 2$ matrix identity
$H \sigma_x H = \sigma_z$ for $H=\frac{1}{\sqrt{2}} \sum_{j,k=0}^{1}
(-1)^{jk} \ket{k}\bra{j}$.  Thus, the above also realizes
$\wedge_1(V)$ in roughly $2d^2$ controlled-$\pi$ phase gates.
This is half the roughly $4d^2$ gates of earlier
work \cite{selectionQR}, even after including
the arbitrary relative phase $\mbox{e}^{i \theta}$ allowed there.

\section{Qudit control without ancillas}
\label{sec:new_control}

In this section we simulate a $\wedge_{n-1}(V)$ gate 
for $V\in U(d)$ using  $O[(n+1)^{\log_2d+2}]$ singly-controlled 
one qudit gates without ancilla.  The method parallels the techniques 
used in Ref. \cite{BarencoEtAl:95} for universal computation with qubits.  

First we decompose a $\wedge_{n-1}(V)$ gate using a 
sequence of gates with a smaller number of controls.  
As a first step, notice that
\begin{equation}
\label{eq:decomp}
\wedge_{n-1}(V)\ =\ \wedge_{n-2}(X_{n-1})[\wedge_{n-2}(\INC)
\wedge_{1}(X_{n-1}^{\dagger})]^{d-1}\wedge_{n-2}(\INC)
\wedge_{1}(X_{n-1}^{d-1}), \end{equation}
where $X_{n-1}=V^{1/d}$. 
For example, for $n=7$, 
we have the following circuit:
\begin{equation}
\label{eq:many_controls}
\Qcircuit @R=1.0em @C=1.0em  {
& \ctrl{1} \qw & \qw & & & \qw & \ctrl{1} & \qw & \ctrl{1} & 
\qw & \ctrl{1}& \qw & \ldots & & \ctrl{2} & \qw & \\
& \ctrl{1} \qw & \qw & & & \qw & \ctrl{1} & \qw & \ctrl{1} & \qw 
& \ctrl{1}& \qw & \ldots & &
\ctrl{2} & \qw & \\
& \ctrl{1} \qw & \qw & & & \qw & \ctrl{1} & \qw & \ctrl{1} 
& \qw & \ctrl{1} & \qw& \ldots & &
\ctrl{2} & \qw & \\
& \ctrl{1} \qw & \qw & & & \qw & \ctrl{1} & \qw & \ctrl{1} & \qw 
& \ctrl{1} & \qw& \ldots & &
\ctrl{2} & \qw & \\
& \ctrl{1} \qw & \qw & & & \qw & \ctrl{1} & \qw & \ctrl{1} 
& \qw & \ctrl{1} & \qw& \ldots & &
\ctrl{2} & \qw & \\
& \ctrl{1} \qw & \qw & \push{\cong} &  & \ctrl{1} & \targ & \ctrl{1} 
& \targ &
\ctrl{1} & \targ & \qw& \ldots & & \qw & \qw & \\
& \gate{V} \qw & \qw & & & \gate{X_{n-1}^{d-1}} & \qw & \gate{X_{n-1}^\dagger} & \qw
& \gate{X_{n-1}^\dagger} & \qw & \qw& \ldots & & \gate{X_{n-1}} & \qw
}
\end{equation}
All control operations are conditioned on the control qudits being 
in state $\ket{d-1}$.  
The circuit is designed to cycle over each possible dit value of the control qudit in the 
$\wedge_{1}(X_{n-1})$ gates.
The entire construction then follows by recursive application of 
Equation \ref{eq:decomp} to the last gate.
In theory, this construction
is an exact implementation of $\wedge_{n-1}(V)$.  Yet in practice,
the sequence of matrices $X_j$ obtained by taking the $d$-th root 
of $X_{j+1}$ (with $X_{n}=V$) quickly converges to the identity
matrix as $j$ decreases.  Hence, an approximate implementation results if
the recursion is terminated early. 

As an example of Equation \ref{eq:many_controls}, consider the
generalized Toffoli gate $\wedge_2(\INC)$.  This breaks into
$(d+1)$ variants of singly-controlled $\wedge_1(W)$ gates
along with $d$ extra \CINC \ gates.  Hence
$(d+1)d+d$ \CINC \ gates along with $(d+1)d$ $\CINC^{-1}$ gates
and sundry gates from $U(d)^{\otimes n}$ suffice to emulate
$\wedge_2(\INC)$.

Note that the size of the circuit for $\wedge_{n-2}(\INC)$
that is analogous to the above
grows exponentially in $n$.  However, it is possible 
to simulate $\wedge_{n-2}(\INC)$ more efficiently using 
a sequence of $\wedge_{\lceil(n-1)/2\rceil}(\INC)$ and 
$\wedge_{\lfloor(n-1)/2\rfloor}(\INC)$ gates, proceeding recursively down to
$\wedge_2(\INC)$.  The argument is 
analogous to that used for qubits in Lemma 7.3 in 
Ref. \cite{BarencoEtAl:95} for $n\geq 5$.  The following 
circuit illustrates the method for $n=7$:
\begin{equation}
\label{eq:k_CINC}
\Qcircuit @R=1.0em @C=1.0em  {
& \ctrl{1} \qw & \qw & & & \ctrl{1} & \qw & \ctrl{1} & \qw & \qw  & \ldots & &
 \qw &  \ctrl{1} & \qw & \qw \\
& \ctrl{1} \qw & \qw & &  & \ctrl{1} & \qw & \ctrl{1} & \qw & \qw &  \ldots & &
 \qw &  \ctrl{1} & \qw & \qw\\
& \ctrl{1} \qw & \qw & &  & \ctrl{3} & \qw & \ctrl{3} & \qw & \qw &  \ldots & &
 \qw &  \ctrl{3} & \qw & \qw \\
& \ctrl{1} \qw & \qw & &  & \qw & \ctrl{1} & \qw & \ctrl{1} & \qw & \ldots & &
 \qw & \qw & \ctrl{1} & \qw \\
& \ctrl{2} \qw & \qw & &  & \qw & \ctrl{1} & \qw & \ctrl{1} & \qw & \ldots & &
 \qw & \qw & \ctrl{1} & \qw \\
& \qw & \qw & \push{\cong} &  & \targ & \ctrl{1} & \targ & \ctrl{1} &
 \qw& \ldots & & \qw &  \targ & \ctrl{1} & \qw  \\
& \targ \qw & \qw & & & \qw & \targ & \qw & \targ
& \qw & \ldots  & & \qw & \qw & \targ & \qw 
}
\end{equation}
Ignoring which qudits are controlled or targeted,
the circuit sequence is 
$\wedge_{n-2}(\INC)=[\wedge_{\lfloor(n-1)/2\rfloor}(\INC)\\
\wedge_{\lceil(n-1)/2\rceil}(\INC)]^{d}$.   

For the remainder of this section, we use a tilde to distinguish a count
for $\CINC^{-1}$ from a $\CINC$ count.  Thus, we let
$b_{n-2}$ be the total number of $\CINC$ gates required to emulate
$\wedge_{n-2}(\INC)$, and
$\tilde{b}_{n-2}$ be the similar count for $\CINC^{-1}$.
For Circuit \ref{eq:k_CINC},
\begin{equation}
\begin{array}{lcl}
b_{n-2} & = & d(b_{\lceil (n-1)/2 \rceil} + b_{\lfloor (n-1)/2 \rfloor}), \\
\tilde{b}_{n-2} & = & d(\tilde{b}_{\lceil (n-1)/2 \rceil} + 
\tilde{b}_{\lfloor (n-1)/2 \rfloor}). \\
\end{array}
\end{equation}
A quick induction shows that each sequence is increasing, and thus
$b_{n-2} \leq 2d b_{\lceil (n-1)/2 \rceil}$ 
and $\tilde{b}_{n-2} \leq 2d \tilde{b}_{\lceil (n-1)/2 \rceil}$.
Moreover, by the analysis of $\wedge_2(\INC)$ above 
$b_2=d^2+2d$ and $\tilde{b}_2=d^2+d$.  Recalling
$(\log_d n)(\log_2 d) = \log_2 n$, we obtain the following:
\begin{equation}
\begin{array}{lclcl}
b_{n-2} & \leq & (d^2+2d)(2d)(2d)^{\log_2 n} & = & 
(d^2+2d)(2d)n^{1+\log_2 d}, \\
\tilde{b}_{n-2} & \leq & (d^2+d)(2d)(2d)^{\log_2 n} & = & 
(d^2+d)(2d)n^{1+\log_2 d}. \\
\end{array}
\end{equation}
Note that these counts assume that the emulation of $\wedge_{n-2}(\INC)$ 
is done on a system with $n$ qudits.   Combining this circuit with 
Circuit \ref{eq:many_controls} allows for an ancilla-free implementation 
of $\wedge_{n-1}(V)$.

Thus, let $c_{n-1}$ be the number of \CINC \ gates required to emulate
$\wedge_{n-1}(V)$, not counting an additional $\tilde{c}_{n-1}$
$\CINC^{-1}$ gates.  Using Circuit \ref{eq:many_controls},
\begin{equation}
\begin{array}{lcl}
c_{n-1} & = & db_{n-2} + c_{n-2} + d^2, \\
\tilde{c}_{n-1} & = & d \tilde{b}_{n-2} + \tilde{c}_{n-2}+d^2. \\
\end{array}
\end{equation}
We may then overestimate $c_{n-1}$ and $\tilde{c}_{n-1}$
using integral comparison and $c_2=d^2+2 d$, $\tilde{c}_2=d^2+d$,
obtaining
\begin{equation}
\begin{array}{lcl}
c_{n-1} & = & d \big( \sum_{j=2}^{n-2} b_j \big) + c_2 + (n-3)d^2 \\
\\
& \leq & d[(d^2+2d)(2d)] \int_4^{n+1} t^{1+\log_2 d} \; dt + 2 d+ (n-2)d^2 \\
\\
& = & \frac{(2d^2)(d^2+2d)}{2+\log_2d} \big[(n+1)^{2+\log_2d} -4 d^2\big]
+ (n-2)d^2+2 d .\\
\end{array}
\end{equation}
We may similarly overestimate $\tilde{c}_{n-1}$:
\begin{equation}
\tilde{c}_{n-1} \ \leq \  
\frac{(2d^2)(d^2+d)}{2+\log_2d} \big[(n+1)^{2+\log_2d} -4 d^2\big]
+ (n-2)d^2+d.
\end{equation}
Hence $c_{n-1}$, $\tilde{c}_{n-1}$ are both bounded by
$O[(n+1)^{2+\log_2d}]$.  This can be used to
show that the earlier spectral algorithm \cite{sharpqudit}
is asymptotically optimal even when ancilla qudits are absent.

If we 
disallow $\CINC^{-1}$ and rather emulate $\CINC^{-1}=\CINC^{d-1}$,
then the overall \CINC \ count for $\wedge_{n-1}(V)$ would be
$c_{n-1}+(d-1)\tilde{c}_{n-1}$.  Note that if the gate library contains 
the two qudit gate
$\wedge_{1}(\sigma_x \oplus I_{d-2})$ rather than $\CINC$, a 
na{\"i}ve application of the above argument would imply a 
linear overhead with a factor of $d-1$.  However 
Circuits \ref{eq:many_controls}
and \ref{eq:k_CINC} can be adapted 
by replacing the $\wedge_k(\INC)$ gates with gates locally 
equivalent to $\wedge_{1}(\sigma_x \oplus I_{d-2})$, resulting 
in a smaller overhead.

\section{Asymptotically optimal qudit state synthesis}
\label{sec:algorithm}

State-synthesis is an important
problem in quantum circuit design \cite{Deutsch_state,Knill_state}.
This section expands upon the earlier account \cite{sharpqudit} of 
an asymptotically optimal state synthesis circuit for qudits.
The earlier circuit used only $O(d^n)$ two-qudit gates, while
a dimension-based argument \cite{sharpqudit} shows that no fewer
($\Omega(d^n)$) gates may achieve qudit state synthesis.
There are two extensions in the present account:
\begin{itemize}
\item  We introduce the $\clubsuit$-sequence, a combinatorial gadget
that organizes the order in which amplitudes are zeroed while (de)constructing
the target state.
\item  Using the $\clubsuit$-sequence, we prove that the
state synthesis algorithm functions as asserted.
\end{itemize}
The two-qudit gates are in fact all $\wedge_1(V)$ for $V$ a one-qudit
Householder reflection.  Hence, earlier sections of the present work
further improve the previous circuit.

Recall from the introduction that we prefer to build $W$ with
$W \ket{\psi}=\ket{0}$ rather than building $U$ with $U \ket{0}=\ket{\psi}$. 
We do this by constructing a sequence of factors which introduce more zeros
into the partially zeroed state.
The ordering established here by the
$\clubsuit$-sequence may be replaced by
Gray code ordering \cite{Vartiainen:04} in the case $d=2$.

\subsection{One-qudit Householder reflections}
\label{sec:house}

Earlier universal $d=2$ circuits
\cite{BarencoEtAl:95} relied on a $QR$ factorization to write
any unitary $U$ as a product of \emph{Givens rotations}, realized
in the circuit as $k$-controlled unitaries \cite{Cybenko:01}.  
In the multi-level case, we instead use 
\emph{Householder reflections} \cite[\S5.1]{gvl}.
Thus, suppose $\ket{\psi} \in \mathcal{H}(1,d)$, perhaps not
normalized.  Householder reflections solve the one-qudit case of
the inverse state-synthesis problem.  Suppose
\begin{equation}
\label{eq:eta_equation}
\left\{
\begin{array}{lcl}
\ket{\eta} & = & \ket{\psi}-\sqrt{\langle \psi | \psi \rangle}
\frac{ \langle    0 | \psi \rangle} {\big| \langle     0| \psi \rangle \big| }
\ket{0}, \\
W & = & I_d - (2/\langle \eta | \eta \rangle) \; \ket{\eta}\bra{\eta}. \\
\end{array}
\right.
\end{equation}
Then $W \ket{{\psi}}$ is a multiple of $\ket{0}$.  Geometrically,
$W$ is that unitary matrix which reflects across a plane lying
between $\ket{0}$ and $\ket{\psi}$.

\subsection{Inserting zeroes using Householders in $\clubsuit$-sequence order}
\label{sec:clubsuit}

The $n$-qudit techniques require a bit more notation.  Any term of the
$\clubsuit$-sequence describes a particular instantiation of a
$\wedge_k(V)$ gate, controlled on certain lines determined by the letters
with target determined by the first $\clubsuit$.  We 
next expand the controlled
operator notation so as to precisely describe how to extract a control
from such a term.

\vbox{
\begin{definition}\cite{sharpqudit}
[Controlled one-qudit operator $\wedge(C,V)$]
\label{def:uniform_control}
Let $V$ be a $d \times d$ unitary matrix, i.e. a one-qudit operator. 
Let $C=[ C_1 C_2 \ldots C_n]$ 
be a length-$n$ \emph{control word} 
composed of letters from the alphabet
$\{0,1,\ldots,d-1\} \sqcup \{ \ast \} \sqcup \{T\}$,
with exactly one letter in the word being $T$.
By $\#C$ we mean the number of letters in the word
with numeric values (i.e., the number of controls.) 
The set of control qudits is the corresponding
subset of $\{1,2,\ldots,n\}$ denoting the positions of numeric
values in the word.  A control word \emph{matches} an $n$-dit string
if each numeric value matches.
Then the controlled one-qudit operator $\wedge(C,V)$
is the $n$-qudit operator that applies $V$ to the qudit specified
by the position of $T$
iff the control word matches the data state's $n$-dit string.
More precisely,
in the case when $C_n=T$, then
\begin{equation}
\wedge( [C_1 C_2 \ldots C_{n-1} T],V) 
\ket{c_1 c_2\ldots c_n} \ = \ 
\left\{
\begin{array}{rl}
\ket{c_1 \ldots c_{n-1}} \otimes V\ket{c_n}, & c_j = C_j 
\mbox{ or } C_j=\ast, \ 1 \leq j \leq n-1 \\
\ket{c_1 \ldots c_{n-1} c_n}, & \mbox{otherwise} \\
\end{array}
\right.
\end{equation}
Alternatively, if $C_j=T$ ($j < n$,) 
we consider the unitary (permutation)
operator $\chi_j^n$ that swaps qudits $j$ and $n$.  Thus, 
$\chi_j^n \ket{d_1 d_2 \ldots d_n} =
\ket{d_1 d_2 \ldots d_{j-1} d_n d_{j+1} \ldots d_{n-1} d_j}$.
Control on a word
$C= [ C_1 C_2 \ldots C_{j-1} T C_{j+1} \ldots C_n ]$,
is then given by $\wedge(C,V)= \chi_j^n \wedge(\tilde{C},V)
\chi_j^n$ for
$\tilde{C}=[C_1 C_2 \ldots C_{j-1} C_n C_{j+1} \ldots C_{n-1} T]$.
\end{definition}  
}

In our particular state synthesis algorithm, we can factor $W$ so that
$\prod_{k=1}^{p} \wedge[C(p-k+1),V({p-k+1})] \; \ket{\psi}
\ = \ \ket{0}$ with all $\# C(k) \leq 1$ and $p=(d^n-1)/(d-1)$.
Since each $\# C(k) \leq 1$, each controlled operation is in fact
a two-qudit gate.  The circuit layout depends on the $\clubsuit$-sequence,
defined in Algorithm 1 and illustrated
in Table \ref{fig:clubsuit}.

\begin{table}[hb]

\centerline{\footnotesize
\begin{tabular}{||c|l||}
\hline 
$n$ & $\clubsuit$-sequence, $d=3$ \\
\hline
\hline
$1$ & $\clubsuit$ \\
\hline
$2$ & $0 \clubsuit$, $1 \clubsuit$, $2 \clubsuit$, $\clubsuit \clubsuit$ \\
\hline
$3$ & $0 0 \clubsuit$, $01\clubsuit$, $02 \clubsuit$, $0 \clubsuit \clubsuit$,
$10 \clubsuit$, $11 \clubsuit$, $12 \clubsuit$, $1 \clubsuit \clubsuit$,
$2 0 \clubsuit$, $2 1 \clubsuit$, $2 2 \clubsuit$, $2 \clubsuit \clubsuit$,
$\clubsuit \clubsuit \clubsuit$ \\
\hline
$4$ & $000 \clubsuit$, $001\clubsuit$, $002 \clubsuit$, 
$00 \clubsuit \clubsuit$,
$010 \clubsuit$, $011 \clubsuit$, $012 \clubsuit$, $01 \clubsuit \clubsuit$,
$020 \clubsuit$, $021 \clubsuit$, $022 \clubsuit$, $02 \clubsuit \clubsuit$,
$0\clubsuit \clubsuit \clubsuit$ \\
& $100 \clubsuit$, $101\clubsuit$, $102 \clubsuit$, 
$10 \clubsuit \clubsuit$,
$110 \clubsuit$, $111 \clubsuit$, $112 \clubsuit$, $11 \clubsuit \clubsuit$,
$120 \clubsuit$, $121 \clubsuit$, $122 \clubsuit$, $12 \clubsuit \clubsuit$,
$1\clubsuit \clubsuit \clubsuit$ \\
& $200 \clubsuit$, $201\clubsuit$, $202 \clubsuit$, 
$20 \clubsuit \clubsuit$,
$210 \clubsuit$, $211 \clubsuit$, $212 \clubsuit$, $21 \clubsuit \clubsuit$,
$220 \clubsuit$, $221 \clubsuit$, $222 \clubsuit$, $22 \clubsuit \clubsuit$,
$2\clubsuit \clubsuit \clubsuit$, $\clubsuit \clubsuit \clubsuit \clubsuit$ \\
\hline
\end{tabular}}
\caption{\label{fig:clubsuit}
Sample $\clubsuit$-sequences for $d=3$, i.e. qutrits.}
\end{table}

\bigskip
 
\vbox{

\noindent
{\bf \hrule}
\smallskip

\noindent
Algorithm 1:  $\{ s_{1},\dots,s_p\}$ 
= {\bf Make-$\clubsuit$-sequence}$(d,n)$

\smallskip

\noindent
{\hrule}

\smallskip

\begin{tabular}{l}
\%  {\em We return a sequence of $p=(d^n-1)/(d-1)$ terms, with
$n$ letters each,} \\
\% {\em drawn from the alphabet $\{ 0,1,\dots,d-1,\clubsuit \}$.} \\
Let $\{ \tilde{s}_{j}\}_{j=1}^{\tilde{p}}$ =  
         {\bf Make-$\clubsuit$-sequence} ($d$,$n-1$). \\
{\bf for } $q=0,1,\dots,d-1$ {\bf do} \\
\quad The next $(d^{n-1}-1)/(d-1)$ terms of the sequence
are formed by prefixing the letter $q$ to each \\
\quad term
of the sequence $\{ \tilde{s}_{j}\}$.\\
{\bf end for} \\
The final term of the sequence is $\clubsuit^n$. \\
\end{tabular}

\noindent
{\hrule}
}

\bigskip

The number of elements in the sequence, $(d^n-1)/(d-1)$,
equals the number of uncontrolled or singly-controlled 
one-qudit operators in our state-synthesis circuit.  
To produce the circuit,
it suffices to describe how to extract the control word $C$ from a
term $t$ of the $\clubsuit$-sequence and how to determine $V$ from
the term and $\ket{\psi_j}$, where 
$\ket{\psi_j}=\prod_{k=1}^{j-1} \wedge[C(p-k+1),V({p-k+1})] \; \ket{\psi}$
is the partial product, as shown in the following algorithm.

\bigskip
 
\vbox{
\noindent
{\bf \hrule}

\smallskip

\noindent
Algorithm 2:  \ \ $\wedge(C,V)$ = {\bf Single-$\clubsuit$Householder}
($\clubsuit$ term $t=t_1t_2 \ldots t_n$, \,  $n$-qudit state $\ket{\psi_j})$

\noindent
{\hrule}

\smallskip

\begin{tabular}{l}
Initialize $C = \ast \ast \cdots \ast$ \\
\% {\em Set the target:} \\
Let $\ell$ be the index of the leftmost $\clubsuit$
      and set $C_{\ell} = T$. \\
\% {\em Set a single control if needed:} \\
{\bf if} $t$ contains numeric values greater than 0, \\
\quad Let $q$ be the index of the rightmost such value
      and set $C_q = t_q$. \\
{\bf end if} \\
Given $\ket{\psi_j}=\sum_{k=0}^{d^n-1} \bra{k} \psi_j \rangle \ket{k}$,
form a one-qudit state $\ket{\varphi}=\sum_{k=0}^{d-1} \bra{t_1 t_2 \ldots
t_{\ell-1} k 0 0 \ldots 0} \psi_j \rangle \ket{k}$. \\
Form $V$ as a one-qudit Householder such that  $V\ket{\varphi}=\ket{0}$. \\
\end{tabular}

\noindent
{\hrule}
}

\bigskip

\begin{figure}[htbp]
\vspace*{13pt}
\[
\Qcircuit @R 0.75em @C 0.75em {
& \lstick{2}  & \qw & \qw & & \push{\mbox{Line }1} && \push{\ast} \\
& \lstick{1}  &  \controlu \qw & \qw &&  \push{\mbox{Line }2} 
&& \push{1} \\
& \lstick{0}  & \qw \qwx & \qw &&  \push{\mbox{Line }3} && \push{\ast} \\
& \lstick{0} & \qw \qwx & \qw &&  \push{\mbox{Line }4} && \push{\ast} \\
& \lstick{\clubsuit} & \gate{V} \qwx & \qw && \push{\mbox{Line }5} 
&& \push{T} \\
& \lstick{\clubsuit} & \qw & \qw &&  \push{\mbox{Line }6} && \push{\ast} \\
& \lstick{\clubsuit} &  \qw & \qw && \push{\mbox{Line }7} && \push{\ast}
}
\]
\caption{\label{fig:club_to_control}
Producing a $\wedge(C,V)$ given $V$ and a term of the
$\clubsuit$-sequence, here $t = 2100\clubsuit \clubsuit \clubsuit$ for
seven qudits.  The algorithm for producing $C$ places
the $V$-target symbol $T$ on the leftmost club, here line
$5$.  The active control must then be placed on the least
significant line carrying a nonzero prior to line $5$, here the $1$
on line 2.  (A control on lines 3 or 4 would not
prevent the nonzero $\alpha_0$ of $\ket{\psi_j} = \sum_{k=0}^{d^n-1}
\alpha_k \ket{k}$ from creating new nonzero entries in previously
zeroed positions.)  Thus in this case, 
$C = \ast 1 \ast \ast T \ast \ast$.  The $V$ is
chosen to zero all but one $\alpha_k$ for $k=2100\ell 00$.
}
\end{figure}
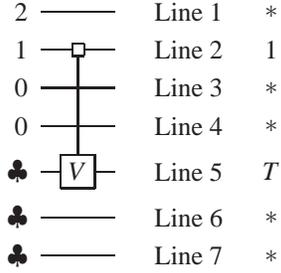

\begin{figure}[htbp]
\centerline{\epsfig{file=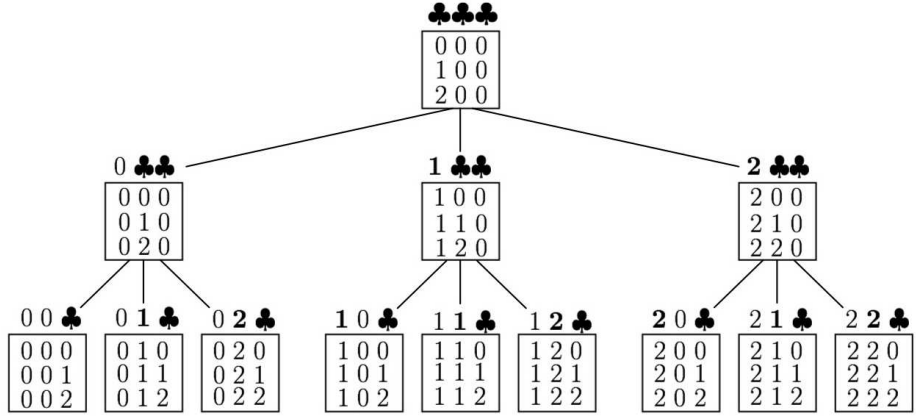, width=12.2cm}} 
\vspace*{13pt}
\caption{
\label{fig:Housegraph}
Using the $\clubsuit$-sequence for $d=3$, $n=3$ to generate
Householder reflections to reduce $\ket{\psi}$ to a multiple of $\ket{0}$.
Each node is labeled by a $\clubsuit$-term
and represents a Householder reflection $\wedge(C,V)$.
The control is indicated by the boldface entry in the label.
As the tree is traversed in a depth-first search, each node indicates a
$\wedge(C,V)$ that zeroes the components of the
last two indices in each node using the component of the top entry.
See also Figure 1 of \cite{sharpqudit}.}
\end{figure}

Figure \ref{fig:club_to_control} displays the type of gate
produced from the output $C$ and $V$ from 
the algorithm {\bf Single-$\clubsuit$Householder}.
Figure \ref{fig:Housegraph} illustrates the order in which these
$\wedge(C,V)$
reflections are generated if we iterate over the
$\clubsuit$-sequence.  Each node of the tree
is labeled by a $\clubsuit$-term
and represents a Householder reflection
defined by three elements of $\ket{\psi}$, whose indices
are indicated in the node.  
The reflection zeroes all but the the first of these three elements.  
The reflections
are applied by traversing the graph in depth-first order, left to
right.

\subsection{Householder circuits for state synthesis}

We will make use of state synthesis for
$\ket{\psi} \mapsto \sqrt{\bra{\psi} \psi \rangle} \ket{0}$
but also for
$\ket{\psi} \mapsto \sqrt{\bra{\psi} \psi \rangle} \ket{m   }$
for any $m   =d_1 d_2 \ldots d_n$.  
We adapt our
construction for a collapse onto $\ket{0}$ into an algorithm
for collapse onto $\ket{m   }$.  
The idea is to permute the elements to put $m   $ in position $0$, 
apply a {\bf Single-$\clubsuit$Householder} sequence,
and then permute back.

Let $m   =d_1d_2d_3\ldots d_n$ be a $d$-ary expansion of some $m   $,
$0 \leq m    \leq d^n-1$.  Then $\ket{m   }=\otimes_{k=1}^n \INC^{d_k} \ket{0}$.
Further, for a generic control word $C$, define 
a new $m   $-dependent control word $\tilde{C}$ by
\begin{equation}
\tilde{C}_k =
\left\{
\begin{array}{rr}
\ast, & C_k = \ast \\
T, & C_k = T \\
(C_k+d_k) \mbox{mod } d, & 
C_k \in \{ 0,1,\dots,d-1\}      \\
\end{array}
\right.
\end{equation}
Suppose also that $C_m    = T$.
Then noting that $(\oplus m   )^\dagger = \oplus (d-m   )$, we have
the similarity relation
\begin{equation}
\label{eq:flip_plus}
[\otimes_{k=1}^n \INC^{d_k}] \wedge(C,V) [\otimes_{k=1}^n
\INC^{d-d_k}] \ = \ 
\wedge[\tilde{C}, (\oplus d_m   ) V (\oplus d-d_m )].
\end{equation}
This is the basis for the algorithm for state synthesis.

\bigskip

\vbox{
\noindent
{\hrule}
\noindent
Algorithm 3:  \ \ $\wedge(C,V)$ = {\bf $\clubsuit$Householder}
$(\ket{\psi},m   ,d,n)$

\noindent
{\hrule}

\smallskip

\begin{tabular}{l}
\% {\em Reduce $\ket{\psi}$ onto $\ket{m   }$.} \\

Let $m   =d_1d_2\ldots d_n$. \\
Compute $\ket{\varphi}
\ =\ (\otimes_{q=1}^n \INC^{d-d_q})\; \ket{\psi}$. \\
Produce a sequence of controlled one-qudit operators so that \\
\quad $\prod_{k=1}^p \wedge[C(p-k+1),V(p-k+1)] 
      \ket{\varphi}=\ket{00\ldots 0}$,\\
\quad  using {\bf Single-$\clubsuit$Householder}
      applied to each term of {\bf Make-$\clubsuit$-sequence}$(d,n)$. \\
Compute
      $(\otimes_{q=1}^n \INC^{d_q} )\wedge[C(p-k+1),V(p-k+1)]
      (\otimes_{p=1}^n \INC^{d-d_q})=$ \\
\quad \quad $\wedge[\tilde{C}(p-k+1), \tilde{V}(p-k+1)]$ 
      using Equation \ref{eq:flip_plus}\\
\end{tabular}

\noindent
{\hrule}
}

{\bf $\clubsuit$Householder}
applies the sequence of Householder reflections
generated by {\bf Single-$\clubsuit$Householder}.
The resulting unitary $W$, although not
a Householder reflection itself,
satisfies $W \ket{\psi}=\ket{m   }$, as we prove in the next
subsection.  Moreover, since the circuit contains $O(d^n)$ two-qudit
gates, all of which are reversible, we have also produced an optimal
gate count for the 
state synthesis problem.  Indeed, if we let $U=W^\dagger$, then
we have $U \ket{0}=\ket{\psi}$.  Moreover, if we label $p(n)=(d^n-1)/(d-1)$,
then $U=\prod_{k=1}^{p(n)} \wedge(C(k),V(k)^\dagger)$
costs $p(n)=O(d^n)$ gates.

We postpone applications to \S \ref{sec:applics} and next
prove that Algorithm 3 is correct.  The proof is new and is organized
in terms of the $\clubsuit$-sequence.

\subsection{Proof that $\clubsuit$-Householder achieves $W \ket{\psi}=\ket{m}$}
\label{sec:no_mix}

For simplicity, we take $m    = 0$, neglecting the permutations.
Given $n$, $p(n)=(d^n-1)/(d-1)$ is the number of elements
of the $\clubsuit$-sequence.  
It would suffice to prove (i) that each operator 
$\wedge[C(j),V(j)]$ guarantees $d-1$ new
zeroes in the state $\ket{\psi_{j+1}}$ not guaranteed in $\ket{\psi_j}$ 
and (ii) moreover that $\wedge[C(j),V(j)]$ does not act on 
previously guaranteed zeroes.
The assertion (i) is straightforward and left to the reader;
see Figure~\ref{fig:Housegraph} caption.
However, the second assertion is \emph{false}.  Rather, the 
controlled one-qudit operators do act
on previously zeroed entries, but always replace them with a zero result.
We next make this assertion precise and prove it.

Define the index set $S=\{0,1,\ldots,d^n-1\}$ and
introduce two new sets of dit-strings:
\begin{itemize}
\item  $S_\ast(j)$ is the set of dit-strings
for which the corresponding amplitude of $\ket{\psi_j}$
is not guaranteed zero by some $\wedge[C(k),V(k)]$, 
$k < j$.  
\item  $S[C(j)]$ is the set of dit-strings that match
$C(j)$, per Definition \ref{def:uniform_control}.
\end{itemize}
Also, define $\ell$ to be the index of the target symbol
in $C(j)$: $C(j)_\ell = T$.
Now there is a group action of $\mathbb{Z}/d\mathbb{Z}$ on the index
set $S$ corresponding to addition mod $d$ on the $\ell^{\mbox{th}}$ dit:
\begin{equation}
c \; \bullet_\ell \ c_1 c_2 \ldots c_n \ = \ 
c_1 c_2 \ldots c_{\ell-1} (c_\ell + c \mbox{ mod d}) c_{\ell+1} \ldots c_n .
\end{equation}
Since the operator $V(j)$ is applied to qudit $\ell$, the
amplitudes (components)
of $\ket{\psi_{j+1}}$ are either equal to the corresponding amplitude
of $\ket{\psi_j}$ or else are linear combinations of the
$\ket{\psi_j}$-amplitudes whose indices lie in the $\mathbb{Z}/d\mathbb{Z}$
orbit contained in $S[C(j)]$.  To establish the correctness
of {\bf $\clubsuit$Householder}, we will prove
the following Proposition.

\begin{proposition}
\label{prop:one}
$\ket{\psi_{j+1}}$ has at least $d-1$ more 
guaranteed zero amplitudes than $\ket{\psi_j}$.
\end{proposition}

Since {\bf $\clubsuit$Householder} sets $j = 1,\dots,(d^n-1)/(d-1)$,
this means that the final $\ket{\psi_j}$ has a single nonzero element
corresponding to $\ket{0}$ and state synthesis has been achieved.
We prove this result using three lemmas.

First we write $S_\ast(j)$ as the union of the three sets 
$R_1(j)$, $R_2(j)$, and $R_3(j)$ which we now define.

\vbox{
\begin{definition}
Suppose the $j^{\mbox{th}}$ term of the $\clubsuit$-sequence is given
by $c_1 c_2 \ldots c_{\ell-1} \clubsuit \ldots \clubsuit$.
We have $C(j)$ the corresponding control word,
with $C(j)_\ell=T$.
Consider the following three sets, noting $R_1(j)$ may be empty.
\begin{equation}
\begin{array}{lcl}
R_1(j) & = & 
\bigsqcup_{q=0}^{\ell-2} \bigg\{ \ c_1 c_2 \ldots c_{q} k 0 0 \cdots 0 \; \;
; \; k < c_{q+1}, k \in \{0,1,\ldots,d-1\} \  \bigg\} \\
R_2(j) & = & 
\bigg\{ \ c_1 \cdots c_{\ell-1} k 0 0 \ldots 0 \; ; \;
k\in \{0,1,\ldots,d-1\} \ \bigg\} \\
R_3(j) & = & 
\bigg\{ \  f_1 \cdots f_{\ell-1} k_\ell k_{\ell+1} \ldots k_n \; ; \;
f_1 f_2 \ldots f_{\ell-1} > c_1 c_2 \cdots c_{\ell-1},
k_\ast \in \{0,1,\ldots,d-1\} \  \bigg\} \\
\end{array} 
\end{equation}
\end{definition}
}

These sets may be interpreted in terms of Figure
\ref{fig:Housegraph}.  Recall the figure recovers the $\clubsuit$-sequence
by doing a depth-first search of the tree.  In this context,
$S_\ast(j)$ is the set of possibly
nonzero components of $\ket{\psi_j}$ at the
$j^{\mbox{th}}$ node.  The subset $R_3(j)$ results from indices that
lie in nodes not yet traversed, loosely
above the present node in the tree or to the right.  
The set $R_2(j)$ is precisely
the set of indices in the current node, node $j$.  The set
$R_1(j)$ is the set of indices of elements that have
been previously used to zero other elements and still might remain
nonzero themselves; 
it is the set of indices of elements
that were always at the top of nodes
already traversed in the depth-first
search.  Thus, $R_1(j)$ is loosely a set of entries within nodes to the
left and perhaps below node $j$.

The first lemma, along with the third,
is used to show that the algorithm does not
harm previously-introduced zeroes.

\begin{lemma}
\label{lem:Sorbit}
Suppose the $\ell^{\mbox{\footnotesize th}}$ letter of $C(j)$ is the target
symbol $T$, and
label $\tilde{S}_\ast(j)=R_1(j) \sqcup R_2(j) \sqcup R_3(j)$.  Then
\begin{equation}
(\mathbb{Z}/d\mathbb{Z}) \; \bullet_\ell \  
\tilde{S}_\ast(j) \cap S[C(j)]  \ \subseteq \  
\tilde{S}_\ast(j) \cap S[C(j)].
\end{equation}
\end{lemma}

\begin{prf}
Due to the choice of a single control on
a dit to the right of position $\ell$ in the appropriate
term of the $\clubsuit$-sequence,
$R_1(j) \cap S[C(j)]= \emptyset$.  On the other hand,
a direct computation verifies
that $(\mathbb{Z}/d\mathbb{Z})\bullet_\ell R_2(j) \subset
R_2(j)$ and also that
$R_2(j) \cap S[C(j)]=R_2(j)$.

Finally, we argue that
$(\mathbb{Z}/d\mathbb{Z})\bullet_\ell R_3(j) \subset R_3(j)$.
However, the following partition is in general nontrivial:
\begin{equation}
\label{eq:R3part}
R_3(j)\ = \ \{R_3(j) \cap S[C(j)] \} \sqcup
\{R_3(j) \cap \big(S-S[C(j)]\big) \} .
\end{equation}
Should $C(j)$ admit no control, we are done.  If not,
let $m<\ell$ be the control qudit.  Then
\begin{equation}
R_3(j) \cap S[C(j)] =  
\bigg\{   f_1 \cdots f_{\ell-1} k_\ell k_{\ell+1} \ldots k_n \; ; \;
{\bf f_m = c_m},
f_1  \ldots f_{\ell-1} > c_1 c_2 \cdots c_{\ell-1},
k_\ast \in \{0,1,\ldots,d-1\}  \bigg\}.
\end{equation}
Hence the $\mathbb{Z}/d\mathbb{Z}$ action respects the partition
of Equation \ref{eq:R3part} as well.
\end{prf}

The second lemma shows that the algorithm produces $d-1$ newly
guaranteed zeroes at each step.

\vbox{
\begin{lemma}
\label{lem:Zlemma}
Let $C(j)$, $\ell$, and $S_\ast(j)$ be as above,
with $C(j)$ resulting from
$c_1 c_2 \ldots c_{\ell-1} \clubsuit \ldots \clubsuit \clubsuit$ of the
$\clubsuit$-sequence.
Let $\mathcal{Z}=
\{c_1 c_2 \ldots c_{\ell-1} k 0 0 \ldots 0 \; ; \; k\in \{1,2,\ldots,d-1\} \cap
\mathbb{Z} \}$ be the elements zeroed by $\wedge(C(j),V(j))$.
Then $R_1(j) \sqcup R_2(j) \sqcup R_3(j)=R_1(j+1) \sqcup R_2(j+1) \sqcup
R_3(j+1) \sqcup \mathcal{Z}$.
\end{lemma}
}

\begin{prf}
We break our argument into two cases based on the value of $c_{\ell-1}$.

\noindent
{\bf Case $c_{\ell-1} < d-1$:}
The $(j+1)^{\mbox{st}}$ term of the $\clubsuit$-sequence is
is given by \hbox{$c_1 c_2 \ldots (c_{\ell-1}+1) 0 0 \ldots 0 \clubsuit$}.
Note that for leaves of the tree, the buffering sequence of zeroes
is vacuous.
\begin{equation}
\begin{array}{lcl}
R_1(j+1) & = & R_1(j) \sqcup R_2(j) -\mathcal{Z}, \\
R_2(j+1) \sqcup R_3(j+1) & = & R_3(j). \\
\end{array}
\end{equation}
 Hence
$R_1(j) \sqcup R_2(j)\sqcup R_3(j) = R_1(j+1) \sqcup R_2(j+1)
\sqcup R_3(j+1) \sqcup \mathcal{Z}$.

\noindent
{\bf Case $c_{\ell-1}=d-1$:}
Suppose instead the $j^{\mbox{th}}$ $\clubsuit$-sequence term is 
$c_1 c_2 \ldots c_{\ell-2} (d-1) \clubsuit \clubsuit \ldots \clubsuit$,
so that the $(j+1)^{\mbox{st}}$ term is
$c_1 c_2 \ldots c_{\ell-2} \clubsuit \clubsuit \clubsuit \ldots \clubsuit$.
We note that 
$\{c_0 c_1 \ldots c_{\ell-2} (d-1) 0 \ldots 0\} 
\in R_2(j) \cap R_2(j+1)$.\footnote{So in the application, the amplitude
(component) of this index is the single amplitude not zeroed by
$\wedge[C(j),V(j)]$, but it is immediately afterwards
zeroed by $\wedge[C(j+1),V(j+1)]$.}
Then
\begin{equation}
\begin{array}{lcl}
R_1(j) & = & R_1(j+1) \sqcup R_2(j+1) -\{ c_0 c_1 \ldots c_{\ell-2} (d-1)
0 \ldots 0\}, \\
R_2(j) & = & \mathcal{Z} \sqcup \{c_0 c_1 \ldots c_{\ell-2} (d-1) 0 \ldots 0\} ,
\\
R_3(j) & = & R_3(j+1). \\
\end{array}
\end{equation}
From the first two, 
$R_1(j) \sqcup R_2(j) = R_1(j+1) \sqcup R_2(j+1) \sqcup \mathcal{Z}$.
Hence $R_1(j) \sqcup R_2(j)\sqcup R_3(j) = R_1(j+1) \sqcup R_2(j+1)
\sqcup R_3(j+1) \sqcup \mathcal{Z}$.
\end{prf}

The third lemma shows that the set we considered in Lemma \ref{lem:Sorbit}
is indeed the set of guaranteed zeros.

\begin{lemma}
\label{lemma:drop_tilde}
$S_\ast(j)= R_1(j) \sqcup R_2(j) \sqcup R_3(j)$ is the set of 
guaranteed zero amplitudes
(components) of a generic $\ket{\psi_j}$.
\end{lemma}

\begin{prf}
The proof is by induction.  For $j=1$, we have
\begin{equation}
R_1(1)=\emptyset, \quad 
R_2(1)=\{00\ldots 0 \ast\}, \quad 
R_3(1)=\{ c_1 c_2 \ldots c_{n-1} \ast \; ; \; \mbox{ some } c_j > 0\}.
\end{equation}
Hence the entire index set $S=S_\ast(1)= R_1(1) \sqcup R_2(1) \sqcup R_3(1)$.

Hence, we suppose by way of induction that
$S_\ast(j)=R_1(j) \sqcup R_2(j) \sqcup R_3(j)$ and attempt to
prove the similar statement for $j+1$.  
Now $\wedge[C(j),V(j)]$ will add new zeroes to the amplitudes
(components) with indices $\mathcal{Z}$ by Lemma \ref{lem:Zlemma}.
On the other hand, $\wedge[C(j),V(j)]$ will not destroy any
zero amplitudes existing in $S_\ast(j)$ 
due to the induction hypothesis and
Lemma \ref{lem:Sorbit}.  Thus
$S_\ast(j+1)=R_1(j+1) \sqcup R_2(j+1) \sqcup R_3(j+1)$.
\end{prf}

\begin{prf}[\ of \ref{prop:one}]
The main result now follows after combining
our three lemmas.
\end{prf}

\section{Unitary synthesis by reduction to triangular form}
\label{sec:triangle}

In this section, we present an asymptotically optimal unitary
circuit not found in \cite{sharpqudit}.  It
leans heavily on the optimal state-synthesis of $\clubsuit$Householder.
Since this state-synthesis circuit can likewise clear any
length $d^n$ vector using fewer than $d^n$ single controls, the
asymptotic is perhaps unsurprising.  Yet the unitary circuit
requires highly-controlled one-qudit unitary
operators when clearing entries near the diagonal.  Optimality
persists since these are used sparingly.  Two themes should
be made clear at the outset:
\begin{itemize}
\item  We process the size $d^n \times d^n$ unitary $V$ in subblocks of
size $d^{n-1} \times d^{n-1}$.
\item  Due to rank considerations, at least one block in each
block-column of size $d^{n}\times d^{n-1}$ must remain full rank throughout.
\end{itemize}
Hence, we cannot carelessly zero subcolumns.  One solution is to
triangularize the $d^{n-1} \times d^{n-1}$ matrices
on the block diagonal, recursively.  Given that strategy,
the counts below show only $O(n^2d^n)$
fully ($n-1$) controlled one-qudit operations appear in the algorithm.
This is allowed when working towards an asymptotic of 
$O(d^{2n})$ gates.

The organization for the algorithm is then as follows.  Processing
(triangularization) of $V$ moves along block-columns of
size $d^n \times d^{n-1}$ from left to right.  In each block-column,
we first triangularize the block $d^{n-1} \times d^{n-1}$ block-diagonal
element, perhaps adding a control on the most significant qudit
to a circuit produced by recursive triangularization.  After this recursion,
we zero the blocks below the block-diagonal element one column
at a time.  For each column $j$, $0 \leq j \leq d^{n-1}-1$, the zeroing
process is to collapse the $d^{n-1} \times 1$ subcolumns onto their
$j^{\mbox{th}}$ entries, again adding a 
control on the most significant
qudit to prevent destroying earlier work.  These subcolumn collapses
produce the bulk of the zeroes and are done using {\bf $\clubsuit$Householder}.
After this, fewer than $d$ entries remain to be zeroed in the column
below the diagonal.  These are eliminated using a controlled
reflection containing $n-1$ controls and targeting the top line.

We now give a formal statement of the algorithm.
We emphasize the addition of controls when
previously generated circuits are incorporated into the universal circuit
(i.e. recursively telescoping control.)

\bigskip

\vbox{
\noindent
{\bf \hrule}

\noindent
Algorithm 4:  {\bf Triangle$(U,d,n)$}

\noindent
{\hrule}

\begin{tabular}{l}
{\bf if } $n=1$ {\bf then} \\
\quad Triangularize $U$ using a $QR$ reduction. \\
{\bf else} \\
\quad Reduce top-left $d^{n-1} \times d^{n-1}$ subblock using
{\bf Triangle$(\ast,d,n-1)$}, (writing output to bottom \\
\quad \ \ $n-1$ circuit lines) \\
\quad {\bf for} $m=0,1,\dots,d-1$ 
{\bf do} \quad \% {\em Block-column iteration} \\
\quad \quad {\bf for} columns $j=md^{n-1},\dots,[(m+1)d^{n-1}-1]$ {\bf do} \\
\quad \quad \quad {\bf for} $\ell=(m+1),\dots,(d-1)$ 
{\bf do} \% {\em Block-row iterate} \\
\quad \quad \quad \quad Use $\clubsuit${\bf Householder} to zero the 
column entries $(m+\ell)d^{n-1},\dots,[(m+\ell+1)d^{n-1}-1]$, \\
\quad \quad \quad \quad \ \ {leaving a nonzero entry}
at $(m+\ell)c_2 \ldots c_n$ for $j=c_1c_2\ldots c_n$ and \\
\quad \quad \quad \quad \ \ {adding $\ket{m+\ell}$- control
on the most significant qudit}. \\
\quad \quad \quad {\bf end for} \\
\quad \quad \quad Clear the remaining nonzero entries below diagonal 
using one $\bigwedge( Tc_2\ldots c_n, V)$. \\
\quad \quad {\bf end for} \quad 
\% {\em All subdiagonal entries zero in block-col} \\
\quad \quad Use {\bf Triangle}$(\ast,d,n-1)$ on the 
$d^{n-1}\times d^{n-1}$ matrix
at the $(m+1)^{\mbox{st}}$ block diagonal \\
\quad \quad \ \ {adding $\ket{m+1}$- control to the
most significant qudit}.
\\
\quad {\bf end for} \\
{\bf end if-else} \\
\end{tabular}

\noindent
{\hrule}
}
\bigskip

To generate a circuit for a unitary operator $U$, we use {\bf Triangle}
to reduce $U$ to
a diagonal operator $W=\sum_{j=0}^{d^n-1}
\mbox{e}^{i \phi_j} \ket{j} \bra{j}$.
Now $V$ and $U=WV$ would be indistinguishable if a von Neumann measurement
$\{ \ket{j}\bra{j} \}_{j=0}^{d^n-1}$ were made after each computation.
However, the diagonal is important if $U$ is a computation corresponding
to a subblock of the circuit of a larger computation with other trailing,
entangling interactions.  In this case, the diagonal unitary 
can be simulated with $d^n$ $\wedge_{n-1}(V)$ gates.  Writing 
$j$ in its d-ary expansion, $j=j_0j_1\ldots j_{n-1}$ we have 
$W=\prod_{j=0}^{d^n-1} \otimes_{k=1}^n \INC_k^{j_k}
\wedge_{n-1}(e^{i\phi_j\ket{d-1}\bra{d-1}})\otimes_{k=1}^n\INC_j^{-j_k}$.  
By the argument in \S \ref{sec:new_control}, the gate count for such a 
simulation is $O[d^n (n-1)^{2+\log_2 d}]$.  This is asymptotically 
irrelevant compared to the lower bound.

\subsection{Counting gates and controls}
\label{sec:counts}

Let $h(n,k)$ be the number of $k$-controls required in the 
{\bf Single-$\clubsuit$Householder} 
reduction of some $\ket{\psi} \in \mathcal{H}(n,d)$.  Then
clearly $h(n,k)=0$ for $k \geq 2$.  Moreover, each
$0$-control results from an element of the $\clubsuit$-sequence of
the form $00\ldots 0 \clubsuit \clubsuit \ldots \clubsuit$, and
there are $n$ such sequences.  Thus,
since the number of elements of the $\clubsuit$-sequence is
$(d^n-1)/(d-1)$, we see that
\begin{equation}
\left\{
\begin{array}{rcr}
h(n,1) & = & (d^n-1)/(d-1) - n \\
h(n,0) & = & n \\
\end{array}
\right.
\end{equation}

We next count controls in the matrix algorithm
{\bf Triangle}.
We break the count into two pieces: $g$ for the work outside
the main diagonal blocks and $f$ for the total work.

Let $g(n,k)$ be the number of $k$-controls applied in 
operations in each column that zero the matrix below the
{block diagonal}; this is the total work in the {\bf for} $j$ loops
of {\bf Triangle}.  
We use
{\bf Single-$\clubsuit$Householder} $d(d-1)d^{n-1}/2$ times
since there are $d(d-1)/2$
blocks of size $d^{n-1} \times d^{n-1}$ below the block diagonal,
and we add a single control to those counted in $h$.
The last statement in the loop is executed $d^{n}-d^{n-1}$ times.
Therefore, letting
$\delta_j^k$ be the Kronecker delta, the counts are  
\begin{equation}
g(n,k) \ = \ \delta_k^{n-1} (d^{n}-d^{n-1})+
\frac{1}{2}d(d-1)d^{n-1} h(n-1,k-1) 
\end{equation}
Supposing $n \geq 3$, then we see that
\begin{equation}
\label{eq:subdiag}
g(n,k) \ = \ 
\left\{
\begin{array}{rr}
d^n-d^{n-1}, & k = n-1 \\
0, & n-1 \leq k \leq 3 \\
\frac{1}{2}d^n(d^{n-1}-1) - \frac{1}{2}d^n(d-1)(n-1), & k=2 \\
\frac{1}{2}d^n(d-1)(n-1), & k=1 \\
0, & k=0 \\
\end{array}
\right.
\end{equation}
Finally, let $f(n,k)$ be the total number of $k$-controlled operations
in the {\bf Triangle} reduction, including the block diagonals.  This
work includes that counted in $g$, plus a recursive call to
{\bf Triangle} before the {\bf for} $m$ loop, plus
$(d-1)$ calls within the $k$ loop, for a total of 
\begin{equation}
\label{eq:recurse}
f(n,k) \ = \ g(n,k) + f(n-1,k) + (d-1)f(n-1,k-1),
\end{equation}
with $f(n,0)=1$ and $f(1,k)=0$ for $n,k>0$.

Using the recursive relation of Equation \ref{eq:recurse} and the
counts of Equation \ref{eq:subdiag},
we next argue that {\bf Triangle} has no more than 
$O(d^{2n})$ controls.  The following lemma is helpful.

\begin{lemma}
\label{lem:gross_overestimate}
For sufficiently large $n$, we have $f(n,k) \leq d^{2n-k+4}$.
\end{lemma}

\begin{prf}
By inspection of Equation \ref{eq:subdiag}, we see that
$g(n,k) \leq (1/2) d^{2n-k+2}$ for all $k$ and $n$ large.  
Now $f(n,0)=1$, which we take as an inductive hypothesis
while supposing $f(n-1,\ell)\leq d^{2n-2-\ell+4}=d^{2n-\ell+2}$.
Thus, using the recursion relation of Equation \ref{eq:recurse},
\begin{equation}
\begin{array}{lcl}
f(n,k) & \leq & \frac{1}{2} d^{2n-k+2} + d^{2n-k+2} + (d-1) d^{2n-k+3} \\
& = & d^{2n-k+4} \; 
\big( \; \frac{1}{2d^2} + \frac{1}{d^2} + 1 - \frac{1}{d} \; \big).
\\
\end{array}
\end{equation}
Now since $d> 3/2$, we must have $\frac{1}{d} > \frac{3}{2d^2}$, whence
an inductive proof of the result.
\end{prf}
 
By the results from \S \ref{sec:new_control}, each $k$-controlled 
single-qudit unitary operator costs $c_k=O[(k+2)^{2+\log_2(d)}]$ 
$\CINC$ and $\CINC^{-1}$ gates without ancillas.  The expected number of 
$\CINC$ gates $\ell_{T}$ for the algorithm {\bf Triangle} is 
then given by the weighted sum for the $k$-control gates in the 
diagonalization and the $d^n$ instances of $n-1$-controlled phase 
gates for emulation of the diagonal:  
\begin{equation}
\begin{array}{lll}
\ell_{T}&=&d^n c_{n-1}+\sum_{k=0}^{n-1} c_k f(n,k)\\
&\leq &2(n+1)^{2+\log_2(d)}d^{n+4}+d^{8+2n} 
\sum_{k=0}^{n-1} d^{-k}k^{2+\lceil\log_2 d\rceil}\\
& \leq & 2(n+1)^{2+\log_2(d)}d^{n+4}+ d^{8+2 n} 
\mathrm{Li}_{-(2+\lceil\log_2d\rceil)}(1/d)\\
&\leq&2(n+1)^{2+\log_2(d)}d^{n+4}+26 d^{8+2 n} .
\end{array}
\end{equation}
In the third line we have used the fact that for the 
Polylogarithm function, $\mathrm{Li}_{-(2+\lceil\log_2d\rceil)}(1/d)
\leq \mathrm{Li}_{-3}(1/2)=26$.

\subsection{Comparison with the spectral algorithm}

\begin{figure}[htbp]
\centerline{\epsfig{file=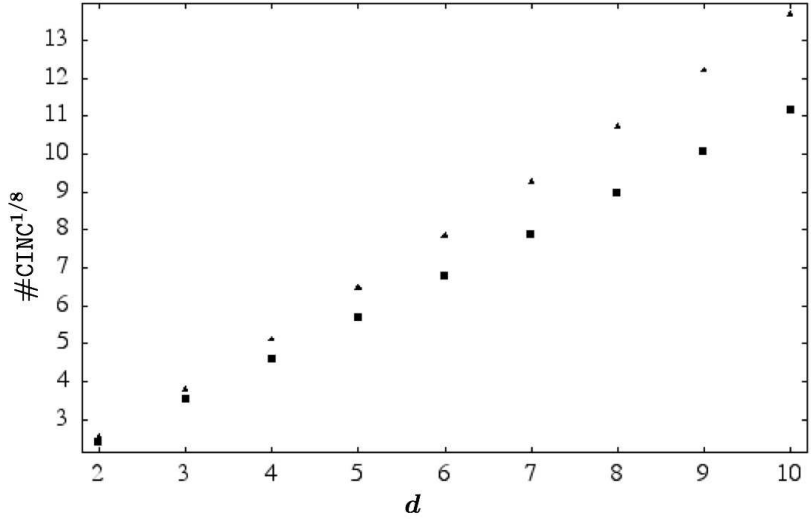, width=12.2cm}} 
\vspace*{13pt}
\caption{
\label{fig:CINCcomp}
Performance comparison of the two algorithms for unitary 
synthesis on $n=4$ qudits as a function of qudit dimension 
$d$.  Triangles (boxes) indicate ${\tt CINC}$ gate counts for 
the {\bf Triangle} (spectral) algorithm.}
\end{figure}

In an earlier work \cite{sharpqudit}, we described an different 
algorithm for unitary synthesis.  That algorithm relied on a 
spectral decomposition of the unitary and was also shown to be 
asymptotically optimal.  For a circuit without ancillas, the 
$\CINC$ gate count $\ell_{S}$ using the spectral algorithm is:
\begin{equation}
\ell_{S}\leq 2 d^{n+1} [ (d^n-1)/(d-1)-n] + (n+1)^{2+\log_2 d}  d^{n+4}
\end{equation}
In Table \ref{fig:control_table_s} the exact gate counts 
resulting from our implementations
for unitary synthesis 
using {\bf Triangle} and the spectral algorithm are tabulated.  The 
result is that for a system with no ancillary resources, the spectral 
algorithm outperforms {\bf Triangle} when the number of qudits $n$ is 
greater than two.  The general $d^{2n}$ scaling for
both is shown in Figure \ref{fig:CINCcomp}.

\begin{table}[hb]
\centerline{\footnotesize
\begin{tabular}{lr||r|r|r|r|r|r|r|r|r||}
\hline 
& $d$ & 2 & 3 & 4 & 5 & 6 & 7 & 8 & 9 & 10 \\
$n$ &   &   &   &   &   &   &   &   &   & \\
\hline
\hline
2 & & {\bf 18} & {\bf 78} & {\bf 220} & {\bf 495} & {\bf 996} 
& {\bf 1 708} & {\bf 2 808} & {\bf 4 365} & {\bf 6 490} \\
 & & {\bf 18} & {\bf 78} & {\bf 220} & {\bf 495} & {\bf 996} 
& {\bf 1 708} & {\bf 2 808} & {\bf 4 365} & {\bf 6 490} \\
\hline 
3 & & 192 & 2 025 & 10 752 & 39 375 & 114 048 & 280 917 & 614 400 
& 1 226 907 & 2 280 000 \\
 & & {\bf 154} & 1 944 & 10 496 & 38 750 & 112 752 & 278 516 & 610 304 
& 1 220 346 & 2 270 000 \\
\hline
4 & & 1 152 & 23 085 & 200 704 & 1 096 875 & 4 447 872 & 14 638 897 
& 41 287 680 & 
103 394 799 & 235 600 000 \\
 & & 1 056 & 22 113 & 195 584 & 1 078 125 & 4 393 440 & 14 504 441 
& 40 992 768 & 
102 804 309 & 234 500 000 \\
\hline
5 & &  5 504 & 223 074 & 3 317 760 & 27 875 000 & 161 523 072 
& 720 717 774 & 2 649 227 264
 & 8 386 138 980 & 23 574 000 000\\
 & & 4 928 & 211 410 & 3 215 360 & 27 312 500 & 159 236 928 
& 713 188 238 & 2 627 993 600
 & 8 332 994 880 & 23 453 000 000\\
\hline
6 & & 23 296 & 1 931 121 & 50 003 968 &  &  & & & & \\
 & & 21 120 & 1 856 763 & 49 070 080 &  &  & & & & \\
\hline
7 & & 92 672 & 16 605 891 & &  & & & & & \\
 & & 84 224 & 16 087 572 &  &  & & & & & \\
\hline
8 & & 353 280 & 141 599 502 &  & & & & & & \\
 & & 324 096 & 138 627 369 &  & & & & & & \\
\hline
9 & & 1 333 248 & 1 224 144 819 & & & & & & & \\
 & & 1 246 208 & 1 209 914 010 & & & & & & & \\
\hline
10 & & 5 025 792 & 10 741 839 786 & & & & & & & \\
 & & 4 786 176 & 10 680 015 483 & & & & & & & \\
\hline
11 & & 19 128 320 &95 432 986 134 & & & & & & & \\
 & & 18 452 480 & 95 147 070 876 & & & & & & & \\
\hline
12 & & 73 515 008 & & & & & & & & \\
 & & 71 639 040 & & & & & & & & \\
\hline
\hline\\
\end{tabular}}
\caption{\label{fig:control_table_s}
Exact gate counts for unitary synthesis without ancillas as a function 
of the number, $n$, and dimension, $d$, of the qudits.  Each cell of the 
table lists the count for $\CINC$ and $\CINC^{-1}$ gates using the 
most efficient of the two algorithms presented in the text.  Boldface 
entries indicate that the {\bf Triangle} algorithm was the 
most efficient, normal face type corresponds to counts using the 
sprectal algorithm.}
\end{table}

There are situations where {\bf Triangle} may be preferred over the 
spectral algorithm.  The later requires a classical 
diagonalization of the unitary $U$ which requires $O(d^{3 n})$ steps.  
For matrices of large size, particularly when 
there are degenerate eigenstates, 
numerical stability can be an issue.  The classical computations involved 
in {\bf Triangle} also scale like $O(d^{3 n})$ but are carried out directly 
in the logical basis of the qudits. 

\section{Two applications of state synthesis}
\label{sec:applics}

A primary motivation for describing state synthesis circuits is
to utilize them as subcircuits for unitary synthesis as in
\S \ref{sec:triangle}.  Yet there are also 
independent applications for the 
state-synth algorithm.  We present two such.

\subsection{Computing expected values}
First, consider the problem of computing the 
expectation value of a Hermitian 
operator $A \in \mathcal{H}(n,d)$ i.e. 
$A \in \mbox{End}[\mathcal{H}(n,d)] \cong \mathbb{C}^{d^n \times d^n}$ 
with $A^{\dagger}=A$.  
For a system in the
possibly mixed state $\rho$ of $n$ qudits, the
the expectation of an operator $A$ is 
\( \langle A \rangle =\mbox{Tr}[A\rho] \). In some cases there does not exist
a physically realistic direct measurement of $A$.  However, one may 
infer the expectation value by a suitably weighted 
set of von Neumann measurements 
as follows. 
By the spectral theorem, any normal operator $A$ 
may be diagonalized by a unitary transformation $U$: $\;$
$A=U^{\dagger}DU$ where 
$D=\sum_{j=0}^{d^n-1}\lambda_j\ket{j}\bra{j}$ and 
$\{\lambda_j\}_{j=0}^{d^n-1}$ are the eigenvalues of $A$.  Then
\begin{equation}
\langle A \rangle \ = \ \mbox{Tr}[A\rho]
\ = \ \mbox{Tr}[DU\rho U^{\dagger}]
\ = \ \sum_{j=0}^{d^n-1}\lambda_j\mbox{\large Tr}\big[
\; \ket{j}\bra{j}U\rho U^{\dagger} \; \big].
\label{expect}
\end{equation}
Hence we may compute $\langle A \rangle$ by performing three steps.
\begin{enumerate}
\item  Prepare $\rho$.
\item  Enact the unitary evolution $U$ on $\rho$.
\item  Perform the computational-basis von Neumann measurement
on the resulting state, extracting all populations of the
basis states $\ket{j}\bra{j}$.
\end{enumerate}

In some instances one may want to know the weight 
of a quantum state on a subspace of the operator $A$, i.e. 
$\langle P_S A P_S\rangle$ where $P_S$ is some projection operator onto 
a subspace $\mathcal{H}_S \subseteq \mathcal{H}(n,d)$.  
In particular, consider the case of a $k$ dimensional subspace diagonal 
in the eigenbasis
$\{\ket{u_j}\}_{j=0}^{d^n-1}$ of $A$.
We wish to compute
$\mbox{Tr}[\sum_{j=1}^k \lambda_j \ket{u_j}\bra{u_j}\rho]$ 
where $k<d^n$ and the eigenvalues of $A$ have been reordered accordingly.  
Then we can rewrite the projection 
$P_SAP_S=\sum_{j=1}^k \lambda_j W(u_j)\ket{j}\bra{j}
W(u_j)^{\dagger}$ where $W(u_j)$ is a unitary extension 
of the mapping $\ket{j}\rightarrow \ket{u_j}$.  The operator 
$W(u_j)$ is the unitary obtained in the state-synth 
algorithm.  The expectation value is then
\begin{equation}
\langle P_SAP_S \rangle\ =\ \sum_{j=1}^k\lambda_j 
\mbox{\large Tr}\big[\; \ket{j}\bra{j}W(u_j)^{\dagger}\rho W(u_j)\; \big].
\end{equation}
The expectation value can be measured as before but now one need 
only implement the state-synth operator $k$ times on each state 
$\rho$ of an ensemble of identically prepared states.

The above argument may in fact be generalized to compute the expectation value of 
any operator A.  First decompose the operator as $A=A_h+A_a$ 
with $A_h=(A+A^{\dagger})/2$ the Hermitian part and $A_a=(A-A^{\dagger})/2$ the anti-Hermitian
part of $A$.  Both $A_h$ and $A_a$ are normal operators and therefore can be diagonalized.  Hence,
the expectation value can be computed by evaluating the weighted sum as per Eq. \ref{expect} and summing.

\subsection{The general state synthesis problem}

Both {\bf Triangle}  and the spectral algorithm
are well adapted to the general state synthesis problem.  This 
problem demands 
synthesizing any unitary extension of the many state  mapping 
$\{\ket{j}\rightarrow \ket{\psi_j} |\  0\leq j\leq \ell\ll d^n\}$
\cite{Knill_state}.  
It is unclear what sorts of applications might arise when the states
are arbitrary, requiring exponentially expensive circuits
to build each.  Nonetheless, less generic
unitaries of this form have been 
used in quantum error correction to encode a few 
logical qudits into many physical qudits \cite{Beth}.

{\bf Triangle} provides one solution to this problem.  Start
with a matrix containing $\ket{\psi_j}$ in its $j$th column,
with  ``don't care" entries in columns after column $\ell$.
Ignore any operations on the ``don't care" entries, and discard
any gates meant to place zeros among them.

The spectral algorithm provides an alternative solution.
Note that the matrix $U$ formed
from the product of the $\ell$ Householder transformations
necessary to reduce the $d^n \times \ell$ matrix $[ \ket{\psi_1} \dots
\ket{\psi_{\ell}}]$ to diagonal form
has $d^n-\ell$ eigenvalues equal to 1, so the spectral algorithm 
needs to build an eigenstate, 
apply a conditional phase to one logical basis ket, 
and unbuild the eigenstate only $\ell$ times.

\section{Conclusions}

This work concerns asymptotically
optimal quantum circuits for qudits.  By asymptotically optimal, we mean
that the circuits require $O(d^{n})$ gates of (no more than) two
qudits for constructing arbitary states and $O(d^{2n})$ gates 
for unitary evolutions.
Contributions of this work are
the following:
\begin{itemize}
\item  We provide the first argument that both asymptotics survive
even when no ancilla (helper) qudits are allowed.
\item  We present the state synthesis circuit in much more detail
than previously published, in particular describing it in terms of
iterates over a $\clubsuit$-sequence which plays a role similar to Gray
codes for bits.  Using the $\clubsuit$ sequence, we provide the
first proof that the
state synthesis circuits actually achieve $U \ket{0}=\ket{\psi}$.
\item  We present {\bf Triangle}, a new asymptotically optimal
quantum circuit for qudit unitaries which is inspired by QR
matrix factorization.  Since it leans more heavily on $QR$ than
on spectral decomposition, the gate parameters of {\bf Triangle} require
less classical pre-processing than the spectral algorithm.  Moreover,
{\bf Triangle} more closely resembles earlier quantum circuit
design techniques \cite{BarencoEtAl:95,Vartiainen:04} than other
asymptotically optimal qudit unitary circuits.
\item \S \ref{sec:one_control} provides an elementary proof that
$\{\CINC\} \sqcup U(d)^{\otimes n}$ is exact-univeral for qudits.
\end{itemize}

Some open questions remain.
The $\wedge_1(V)$ gates are much better than earlier practice but
not provably optimal, as is the case with qubits 
\cite{SongKlappenecker}.  Moreover,
the current best-practice $n$-qubit circuits exploit the 
cosine-sine decomposition (CSD), yet technical difficulties \cite{Werner} with
the tensor product structure make it quite unclear whether
this matrix decomposition is useful for qudits.

{\bf Acknowledgements}\quad DPO received partial 
support from the National Science Foundation under 
Grants CCR-0204084 and CCF-0514213.  GKB was supported in part by a 
grant from DARPA/QUIST.  SSB was supported by a National Research
Council postdoctoral fellowship.
\label{sec:conclusions}

\end{document}